\documentclass{wqcd03}                 

\usepackage{epsfig,amsmath,amssymb}

\usepackage{mathptmx}                 

\newcommand{\eq}{{\, \equiv\, }}
\newcommand{\fr}[1]{
             \frac{#1}}       
\newcommand{\be}{\begin{equation}}
\newcommand{\ee}{\end{equation}}
\newcommand{\bea}{\begin{eqnarray}}
\newcommand{\eea}{\end{eqnarray}}

\newcommand{\chibar}{\overline{\chi}}

\newcommand{\ket}{{\rangle}}

\newcommand{\bra}{{\langle}}
\newcommand{\gc}{\bra\fr{\alpha_s}{\pi}G^2\ket}
\newcommand{\qc}{\bra\,\overline{q}q\,\ket}
\newcommand{\bbar}{B\,-\,\overline{B}}
\newcommand{\ga}{{g_{{\cal A}}}}

\newcommand{\Qp}{Q^{(+)}_v}
\newcommand{\Qm}{Q^{(-)}_v}

\confname{QCD@Work 2003 - International Workshop on QCD, Conversano, Italy, 
14--18 June 2003}

\title{$B$-decays and $B - \overline{B}$ mixing within a
 heavy-light chiral quark model}

\author{Jan O. Eeg}

\address{Department of Physics, University of Oslo, Norway}

\begin{document}

\begin{abstract}
We describe  a recently developed heavy-light chiral quark model and
show how it can be used to calculate decay amplitudes for heavy mesons.
In particular, we discuss  $\bbar$ mixing,  $B \rightarrow D \bar{D}$,
 $B \rightarrow D \eta'$ and the beta term for $D^* \rightarrow D \gamma$.
\end{abstract}

\maketitle


\section{Introduction}

Some $B$-decays
 where  the energy
 release is big  compared to the light
meson masses, for instance
 $B \rightarrow \pi \pi$ and $B \rightarrow K \pi$,
 has been  successfully  described by
{\it QCD factorization} 
 \cite{BBNS,CSachr}.
However, for various  $B$-decays where
 the energy release
is of order 1 GeV or less , QCD factorization is not expected to hold.
The purpose of this presentation is to describe how processes
like  $\bbar$ mixing \cite{ahjoeB},  $B \rightarrow D \bar{D}$
\cite{EFH}, $B \rightarrow D \eta'$ \cite{EHP}, and in addition,
some aspects of $D$-meson decays \cite{EFZ,Beta} can be described
within a recently developed heavy-light chiral quark model 
(HL$\chi$QM) \cite{AHJOE}.

In general, weak  non-leptonic processes  may be described by an effective
Lagrangian which is a linear combination of quark operators where the  
(Wilson) coefficients are
 calculated in perturbation theory.
These quark operators  are bosonized
within the  HL$\chi$QM, 
 where non-factorizable
effects can be incorporated by means of gluon condensates and chiral loops.
The coefficients of the various terms of weak chiral Lagrangians can
then be calculated.
In this way we can make a bridge between the quark and the mesonic sector.

\section{The heavy-light chiral quark model}

Our model is based on the following Lagrangian containing both quark
 and meson fields:
\begin{equation}
{\cal L} =  {\cal L}_{HQEFT} +  {\cal L}_{\chi QM}  +   {\cal L}_{Int} \; ,
\label{totlag}
\end{equation}
where \cite{neu}
\begin{equation}
{\cal L}_{HQEFT} = 
 \overline{Q_{v}} \, i v \cdot D \, Q_{v} 
 + {\cal L}_{HQEFT}^{(1)}
 + {\cal O}(m_Q^{- 2})
\label{LHQEFT}
\end{equation}
is the Lagrangian for heavy quark effective field theory (HQEFT).
The heavy quark field  $Q_v$
annihilates  a heavy quark  with velocity $v$ and mass
$m_Q$. Moreover,  
$D_\mu$ is the covariant derivative containing the gluon field
(eventually also the photon field). 
The first order ($m_Q^{-1}$) term is
\begin{equation}
{\cal L}_{HQEFT}^{(1)} = 
 \frac{1}{2 m_Q}\overline{Q_{v}} \, 
\left( - C_M \frac{g_s}{2}\sigma \cdot G
 \, +   \, (i D_\perp)_{\text{eff}}^2  \right) \, Q_{v} \; ,
\label{LHQEFT1}
\end{equation}
where $\sigma \cdot G = \sigma^{\mu \nu} G^a_{\mu \nu} t^a$, and 
$\sigma^{\mu \nu}= i [\gamma^\mu, \gamma^\nu]/2$.  $G^a_{\mu \nu}$
is the gluonic field tensor, and $t^a$ are the colour matrices ($a=$1,..8). 
This chromo-magnetic term has a factor $C_M$, being one at tree level,
 but slightly modified by perturbative QCD. 
(When the covariant derivative also contains the photon field, there
is also a corresponding magnetic term
$\sim \sigma \cdot F$, where $F^{\mu \nu}$ is the electromagnetic tensor).
 Furthermore,  
$(i D_\perp)_{\text{eff}}^2 =
C_D (i D)^2 - C_K (i v \cdot D)^2 $. At tree level, $C_D = C_K = 1$.
Here, $C_K$ is different 
from one due to perturbative QCD,
while $C_D$ is not modified \cite{GriFa}.

The light quark sector is described by the chiral quark model ($\chi$QM),
having a standard QCD term and a term describing interactions between
quarks and  (Goldstone) mesons \cite{pider}.
Making a flavour rotation of the quark fields $q_L$ and $q_R$
transforming as $SU(3)_L$ and $SU(3)_R$ respectively,
the Lagrangian can be written in terms of quark fields $\chi$
transforming as $SU(3)_V$  triplets (see
refs.\cite{pider,epb,BEF,BEFL}  and references therein):
\begin{equation}
{\cal L}_{\chi QM} =  
\chibar \left[\gamma^\mu (i D_\mu   +    {\cal V}_{\mu}  +  
\gamma_5  {\cal A}_{\mu})    -    m \right]\chi 
  -     \chibar \widetilde{M_q} \chi \;  , 
\label{chqmR}
\end{equation}
where 
$\chi_L  =   \xi^\dagger q_L$ and  $\chi_R  =   \xi q_R$. Here, 
$\xi  =   \exp(i\Pi/f )$, where $\Pi$ is a 3 by 3 matrix containing
 the (would be)  Goldstone octet ($\pi, K, \eta$)
in the standard way, and $f$ is the bare pion decay constant. The quantity
$m$ is the ($SU(3)$ -  invariant) constituent quark mass for light
quarks.
The vector and axial vector fields 
${\cal V}_{\mu}$ and  
${\cal A}_\mu$ are given by:
\begin{equation}
{\cal V}_{\mu} = \frac{i}{2}(\xi^\dagger\partial_\mu\xi
+\xi\partial_\mu\xi^\dagger 
) \; \; ;  \; \;   
{\cal A}_\mu =  -  \frac{i}{2}
(\xi^\dagger\partial_\mu\xi
-\xi\partial_\mu\xi^\dagger) \; ,
\label{defVA}
\end{equation}
and  $\widetilde{M_q}$ defines  the rotated version of the current
light mass matrix  ${\cal M}_q = diag(m_u,m_d,m_s)$:
\bea
\widetilde{M_q} = \widetilde{M}_q^V   +    \widetilde{M}_q^A \gamma_5  \;
 \; ; \quad 
\widetilde{M}_q^{V,A} \, \equiv \, 
\fr{1}{2}(\xi^\dagger {\cal M}_q \xi^\dagger \, \pm \,
 \xi {\cal M}_q^\dagger \xi ) \; .
\label{masst}
\eea
In the light sector, the various pieces of the strong
chiral  Lagrangian
can be obtained by integrating out the constituent quark fields $\chi$,
and these pieces can be written in terms of the  fields ${\cal A}_\mu \, ,
\, \widetilde{M}_q^V$ and $\widetilde{M}_q^A$.
   
In the heavy-light case, the generalization of the
 meson -  quark interactions in the pure light sector  $\chi$QM
is given by the following $SU(3)_V$ 
invariant Lagrangian \cite{AHJOE,barhi,CQM}:
\begin{equation}
{\cal L}_{Int}  =   
 -   G_H \, \left[ \chibar_k \, \overline{H_{v}^{k}} \, Q_{v} \,
  +     \overline{Q_{v}} \, H_{v}^{k} \, \chi_k \right]   \; ,
\label{Int}
\end{equation}
where $G_H$ 
is a coupling constant, and
$H_{v}^{k}$ is the heavy meson field  containing
a spin zero and spin one boson:
\begin{eqnarray}
&H_{v}^{k} & =  P_{+} (P_{\mu}^{k} \gamma^\mu -     
i P_{5}^{k} \gamma_5)\; , \nonumber \\
&\overline{H_{v}^k}
& =  \gamma^0 (H_{v}^k)^\dagger \gamma^0 \; ; \quad 
P_{\pm} =  (1 \pm \gamma \cdot v)/2 \; . 
\label{HMes}
\end{eqnarray}
Here the index $k$ runs over the light quark flavours $u, d, s$.
The fields $P_{5} (P_{\mu})$ annihilates a heavy-light meson with
spin-parity
 $0^{-}(1^-)$,  and velocity $v$. Note that for  antiquarks, the
heavy quark field 
$Q_{v}\eq Q_{v}^{(+)}$ has to be replaced by the heavy $anti$quark field 
$Q_{v}^{(-)}$ in (\ref{LHQEFT}) and  (\ref{Int}).
At the same time, the heavy meson field $H_{v} \eq H_{v}^{(+)}$ in 
(\ref{Int}) and (\ref{HMes}) has to be replaced by the $anti$meson
 field $H_{v}^{(-)}$, and the velocity $v$ is replaced by $(-v)$.

 In our model, the hard gluons are thought to be integrated out and we are
left with soft gluonic degrees of freedom. Emission of such gluons can be
described using  external field techniques  \cite{nov}, and their
effect will be parameterized by vacuum expectation values, 
i.e. the gluon condensate $\gc$. Our model dependent gluon
 condensate contributions are obtained by the replacement
\begin{equation}
g_s^2 G_{\mu \nu}^a G_{\alpha \beta}^b  \; \rightarrow
 \fr{4 \pi^2 \delta^{a b}}{(N_c^2-1)}
 \gc \frac{1}{12} (g_{\mu \alpha} g_{\nu \beta} -  
g_{\mu \beta} g_{\nu \alpha} ) \, .
\label{Gluecond}
\end{equation}
We observe that soft gluons
coupling to a heavy quark is suppressed by $1/m_Q$, since to leading
order the vertex is proportional to $v_\mu v_\nu G^{a\mu\nu}= 0$,
 $v_\mu$ being the heavy quark velocity. 

Note that opposite parity heavy meson states, like the recently
 discovered $D^*$ resonance, can also be incorporated
in the formalism  \cite{DRes}.

\section{Bosonization within the HL$\chi$QM}\label{sec:strong}
 
The interaction term ${\cal{L}}_{Int}$ in (\ref{Int}) can now 
be used to bosonize the model, i.e. 
integrate out the quark fields. This can be done
 in the path integral formalism, 
or in terms of Feynman diagrams by attaching the
 external fields $H_v^{k}, \overline{H_v^{k}},  {\cal{V}}^\mu,
  {\cal{A}}^\mu$ and $\widetilde{M}_q^{V,A}$ of section 2 to quark loops.
Some of the loop integrals will be divergent and have  to be related
to physical parameters, as for the pure
 light sector \cite{pider,epb,BEF,BEFL}.
The strong chiral Lagrangian 
has the following form (see \cite{AHJOE,itchpt} and references therein): 
\bea
{\cal L}_{Str} 
=  \, -  Tr\left[\overline{H_{k}}(iv\cdot D)H_{k}\right]\, + \,
Tr\left[\overline{H_{k}}H_{h}v_\mu {\cal V}^\mu_{hk}
\right]
\nonumber \\ 
- \,g_{\cal A} Tr\left[\overline{H_{k}}H_{h}\gamma_\mu\gamma_5 {\cal
A}^\mu_{hk}\right] + 
 2 \lambda_1 Tr\left[\overline{H_{k}}H_{h} (\widetilde{M}_q^V)_{hk}\right]
+ ...
 \label{LS1}
\eea
where the velocity index on the heavy meson field is suppressed, 
 the ellipses indicate other terms (of higher order, say), and 
$D_\mu$ contains the photon field.
The trace runs over  gamma matrices.

Comparing  the loop integral
 for the  diagrams in figure~\ref{fig:va} with the vector field ${\cal V}_\mu$ 
attached to the light quark, we obtain the following 
identification: 
\begin{equation}
 -  iG_H^2N_c \, (I_{3/2}  +   2mI_2  +  
\fr{i(8-3\pi)}{384N_cm^3}\gc) =   1 \; ,
\label{norm}
\end{equation}
where $I_{3/2}$ and $I_2$ are  
linear -  and logarithmic -  divergent loop integrals
(these have to be interpreted as  the regularized ones).
Note that for the kinetic term in (\ref{LS1}) we obtain the same
relation as (\ref{norm}) due to the relevant Ward identity.

The relation (\ref{norm}) is analogous to  the pure light sector
where the quadratic and logarithmic divergent
integrals are related to $f$ (the bare $f_\pi$) and the quark condensate
\cite{pider,epb,BEF,BEFL}:
\begin{equation}\label{I2}
f^2 =    -  i4m^2N_cI_2 +  \fr{1}{24m^2}\gc \; ,
\end{equation}
\begin{equation}\label{I1}
\qc = -4imN_cI_1-\fr{1}{12m}\gc \; ,
\end{equation}
where $I_1$ is the quadratically divergent loop integral.
As the pure light sector is a part of our model, we have to keep these 
relations in the heavy-light case studied here.

Also from  diagram \ref{fig:va}, with the axial field ${\cal A}_\mu$ attached,
 we obtain a similar identification for the
 axial vector coupling $g_{\cal A}$. Using (\ref{norm}) this can be rewritten:
\begin{figure}[t]
\begin{center}
   \epsfig{file=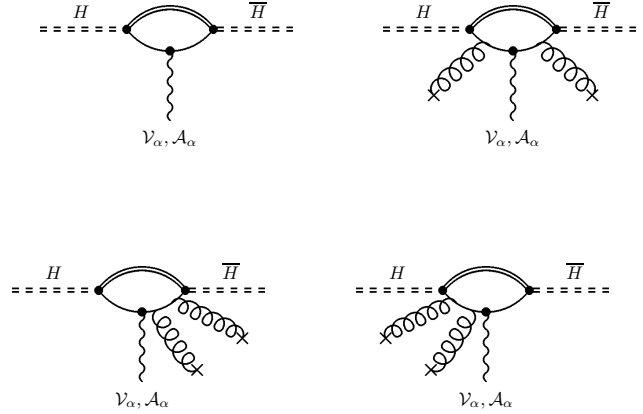,width=8cm}
\caption{Diagrams contributing to
 the lowest order heavy-light chiral Lagrangian.}
\label{fig:va}
\end{center}
\end{figure}
\begin{equation}
   g_{\cal A} = 1  +  \frac{4}{3} i G_H^2N_c \, \left(I_{3/2} 
 - \frac{i m}{16 \pi} \right)  \; .
\label{dga}
\end{equation}
Within a primitive cut-off regularization, $I_{3/2}$ is (in the leading
 approximation) proportional to the cut-off in first power \cite{barhi},
while it is finite in  dimensional regularization. We will keep $I_{3/2}$
as a free parameter to be determined by the physical value of $g_{\cal A}$.

Within HQEFT the weak current will,  below the renormalization scale
 $\mu  =   m_Q \, (=m_b, m_c)$, be modified in the following way:
\begin{equation}
J_k^\alpha
 =  {\chibar}_h \xi^{\dagger}_{hk} \Gamma^\alpha  Q_v     
\, + {\cal O}(m_Q^{-1}) \; ,
\label{modcur}
\end{equation}
where $k$ and $h$ are light flavour indices. 
 The $1/m_Q$ terms contain an extra covariant derivative
\cite{neu} and 
\begin{eqnarray}
\Gamma^\alpha \,\eq\, C_\gamma (\mu )\,\gamma^\alpha\, L \,+\,  
C_v(\mu )\, v^\alpha\, R\; ,
\label{Gamma}
\end{eqnarray}
where $L$ and $R$ are left and right Dirac projection matrices.
The coefficients $C_{\gamma,v}(\mu)$ are determined 
by QCD renormalization for  $\mu < m_Q$ and have been calculated to
NLO.
\begin{figure}[t]
\begin{center}
   \epsfig{file= 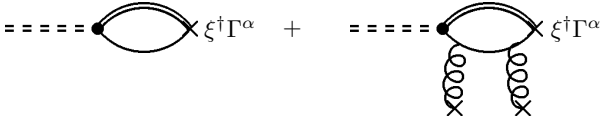,width=8cm}
\caption{Diagrams for bosonization of the heavy-light 
left handed quark current.}
\label{fig:f+}
\end{center}
\end{figure}
 We obtain to zero order in the axial
field ${\cal A}^\mu$ (-or ${\cal O}(p^0)$ in the language of chiral
perturbation theory. See figure~\ref{fig:f+}):
\begin{equation}
J_k^\alpha (0) =  \fr{\alpha_H}{2} Tr\left[\xi^{\dagger}_{hk}\Gamma^\alpha
  H_{vh}\right]
  \; ,\label{J(0)}
\end{equation}
where
\begin{equation}
\alpha_H \eq  -  2iG_HN_c\left(  -  I_1  +  mI_{3/2} +
 \fr{i (3\pi -4)}{384 m^2 N_c}\gc\right) \; .
\label{alphaH}
\end{equation}
We observe that, as this relation involves $I_1$, there is a relation
between $\alpha_H$ and the quark condensate \cite{AHJOE}.

The coupling $\alpha_H$ in (\ref{J(0)}) is related to  the physical
 decay constant $f_H$   by considering (for $H=B,D$):
\bea
\bra 0| \overline{u}\gamma^\alpha \gamma_5 b|H \ket 
 =&   -  2\,\bra 0| J_a^\alpha|H \ket\, 
 &=  iM_Hf_H v^\alpha \; .
\eea
Combining this with (\ref{J(0)}), and adding chiral corrections
and the $1/m_Q$ corrections indicated in
 (\ref{modcur}),  we obtain 
\bea
f_H = \frac{1}{\sqrt{M_H}}\left[\left(C_\gamma (\mu ) + C_v(\mu )\right)  
\alpha_H  + \frac{\eta_Q}{m_Q} + \frac{\eta_\chi}{32\pi^2 f_\pi^2}
 \right]
\; ,
\label{fb}
\eea
where the model dictates us to put $\mu  =   \Lambda_\chi$.
The quantities $\eta_Q$ and  $\eta_\chi$ are given in \cite{AHJOE}.

The gluon condensate can be related to the chromo-magnetic
interaction via the $H-H^*$ mass difference: 
\begin{equation}
\mu_G^2 =  \fr{C_M}{4M_H}
\bra H|\bar{Q_v} \sigma\cdot GQ_v|H\ket \, 
 = \frac{3 m_Q}{2} (M_{H^*} -  M_H)
 \; .   \label{gc2}
\end{equation} 
An explicit calculation of the matrix element 
 in equation (\ref{gc2}) gives
\begin{equation}
\label{mug2}
\mu_G^2  = \fr{(\pi +  2)}{32 m} C_M(\Lambda_\chi) \,  G_H^2 \gc \,
 .
\end{equation}
For further details, see \cite{AHJOE}.

\section{$\bbar$ mixing and heavy quark effective theory}

At quark level, the standard
effective Lagrangian describing $\bbar$ mixing is \cite{EffHam}
\bea
{\cal L}_{eff}^{\Delta B= 2} \, = \,
 - \, \fr{G_F^2}{4\pi^2} M_W^2 \left(V_{tb}^*V_{tq}\right)^2 
\,S_0\left(x_t \right)\,\eta_B \,
b(\mu) \; Q_B \; , 
\label{LB2}
\eea
where  $G_F$ is Fermi's coupling constant, the $V$'s are 
KM factors 
 (for 
which $q=d$ or $s$ for $B_d$ and $B_s$ respectively)
 and $S_0$ is the Inami-Lim function due to
 short distance electroweak
loop effects for the box diagram. 
In our case, $x=x_t \equiv m_t^2/M_W^2$, where 
 $m_t$ is the 
top quark mass. 
The quantity $Q_B \equiv Q(\Delta B=2)$ is a four quark
operator 
\begin{equation}
Q_B= \overline{q_L}\,\gamma^\alpha\, b_L \;
\overline{q_L}\,\gamma_\alpha\, b_L \; ,
\label{QB2}
\end{equation} 
where $q_L$ $(b_L)$
 is the left-handed
projection of the $q$ $(b)$-quark field.
The quantities $\eta_B$ and $b(\mu)$ are calculated in perturbative
quantum chromodynamics (QCD). At the next to leading order (NLO)
analysis it is found that  
$\eta_B= 0.55\pm 0.01\,$ .
At $\mu~=~m_b$ $(=4.8~\text{GeV})$ one has $b(m_b)\simeq 1.56$ 
in the naive dimension regularization scheme (NDR).

The matrix element of the operator $Q_B$ between the meson
 states is parameterized 
by the bag parameter $B_{B_q}$ :
\begin{equation}
\bra B|Q_B|\overline{B}\ket \eq 
\fr{2}{3} f_B^2 M_B^2 B_{B_q}(\mu) \; \, .
\label{matrQ}
\end{equation}
By definition, $B_{B_q} =1 $ within  {\it naive factorization}, also named
vacuum saturation approach (VSA).

In general, the matrix element of  the operator $Q_B$
is dependent on the renormalization scale $\mu$, and thereby $B_{B_q}$
depends on $\mu$. 
 As for $K-\overline{K}$
mixing, one defines a renormalization scale independent quantity
\begin{equation}
\hat{B}_{B_q} \equiv b(\mu) B_{B_q}(\mu) \; .
\label{Bhat}
\end{equation}
Within lattice gauge theory,    values for  $\hat{B}_{B_q}$
 between 1.3 and 1.5 are obtained \cite{latt}.

Running from $\mu = m_b$ down to $\mu = \Lambda_\chi=1\,$ GeV,
 there will appear
more operators. Some stem from the heavy quark expansion itself and some
are  generated by
perturbative QCD effects. The $\Delta B=2$ operator in equation
(\ref{QB2}) for  $\Lambda_\chi < \mu < m_b$ can be written
 \cite{gimenez,mannel} :
\bea
Q_B = \,  C_1 \, Q_1 + C_2 \, Q_2
+\fr{1}{m_b} \sum_i \,h_i X_i
\,+{\cal O}(1/m_b^2) \; \, .
\label{HQ}
\eea
The operator $Q_1$ is $Q_B$ for $b$ replaced by $Q_v^{(\pm)}$,
while $Q_2$ is 
generated within perturbative QCD for $\mu < m_b$.  The operators
 $X_i$ are taking care of $1/m_b$ corrections. The quantities 
$C_1, C_2, h_i$ are Wilson coefficients.
 The  operators are given by  
\bea
&Q_1  &=   2 \;  \overline{q_L}\,\gamma^\mu \,\Qp \; \; 
\overline{q_L}\,\gamma_\mu \,\Qm \, \; , 
\label{Q1}\\
&Q_2  &=   2 \; \overline{q_L}\,v^\mu \,\Qp \; \;
 \overline{q_L}\,v_\mu \,\Qm \; \, ,
\label{Q2}\\
&X_1&= 
 2 \;\overline{q_L}\,iD^\mu \,\Qp 
\;\overline{q_L}\,\gamma_\mu \Qm\ \; + ......
\eea
There are also non-local  operators  constructed as time-ordered
 products of  $Q_{1,2}$ and the first order HQEFT Lagrangian in
(\ref{LHQEFT1}).
The Wilson coefficients $C_1$ and $C_2$ have been calculated to NLO
 \cite{gimenez} and for $\mu = \Lambda_\chi$,
$C_1(\Lambda_\chi)= 1.22$ and $C_2(\Lambda_\chi)= - 0.15$. 
The coefficients $h_i$ have been calculated to leading order (LO)
in \cite{mannel}.

In order to find the matrix element of $Q_{1,2} \,$, one uses the 
following relation between the generators of $SU(3)_c$ ($i,j,l,n$
are colour indices running from 1 to 3):
\begin{equation}
\delta_{i j}\delta_{l n}  =   \fr{1}{N_c} \delta_{i n} \delta_{l j}
 \; +  \; 2 \; t_{i n}^a \; t_{l j}^a \; ,
\label{fierz}
\end{equation}
where $a$ is an index running over the eight gluon charges. This
 means that  by means of a Fierz transformation, the operator $Q_1$
 in (\ref{Q1}) may  also be written in the following way 
(there is a  similar expression for $Q_2$): 
\bea
Q_1  &=&   \fr{2}{N_c}\, \overline{q_L} \,\gamma^\mu\, \Qp\,
\overline{q_L}\, \gamma_\mu\, \Qm \nonumber \\
\,&+&  4\, \overline{q_L}\, t^a\, \gamma^\mu  \,\Qp \,
 \overline{q_L} \,t^a\, \gamma_\mu\, \Qm \, .
\label{Q1Fierz}
\eea

The first (naive) step to calculate the matrix element of a four quark 
operator like $Q_1$ is to insert vacuum states between the two currents.
This vacuum saturation approach (VSA)
means to bosonize the two currents in $Q_1$
(see (\ref{modcur})) and multiply them.

\begin{figure}[t]
\begin{center}
   \epsfig{file=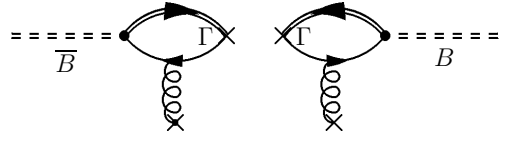,width=7cm}
\caption{Non-factorizable soft gluonic contribution to the
 bag-parameter.
 (Here $\Gamma\eq t^a \,\gamma^\mu\,L$.)}
\label{fig:bbargg}
\end{center}
\end{figure}

 The second operator in (\ref{Q1Fierz}) is genuinely non-factorizable.
In the approximation where only the lowest gluon condensate is
 taken into account, the last term in (\ref{Q1Fierz}) can be written in a 
{\it quasi-factorizable} way by
bosonizating the heavy-light coloured current 
with an extra  colour matrix $t^a$ inserted and with an extra gluon 
emitted as shown in figure~\ref{fig:bbargg}.

We find  the bosonized coloured current:
\bea
\left(\overline{q_L} t^a\,\gamma^\alpha Q_v^{(\pm)}\right)_{1G} 
\;   \longrightarrow \;
 -\fr{G_H g_s}{8}\,G_{\mu\nu}^a \nonumber \\ 
\; \times Tr\left[\xi^\dagger \gamma^\alpha  L \, H^{(\pm)}
\left(\pm i\,I_2\left\{\sigma^{\mu\nu}, \gamma \cdot v\right\}+ \,
\fr{1}{8\pi} \sigma^{\mu\nu} \right)\right] \; ,
\label{1G}
\eea
where $\{ \, , \, \}$ symbolizes an anti-commutator.
The result for the right part of the diagram with $\bar{B}$ replaced by 
$B$  is obtained by 
changing the sign of $v$ and letting ${P_5^{(+)}}\rightarrow {P_5^{(-)}}$
(see the comments below eq. (\ref{HMes})).  
Multiplying the coloured currents, we obtain the non-factorizable
parts of $Q_1$ and $Q_2$ to first order in the gluon condensate
by using eq. (\ref{Gluecond}).

\begin{figure}[t]
\begin{center}
\epsfig{file=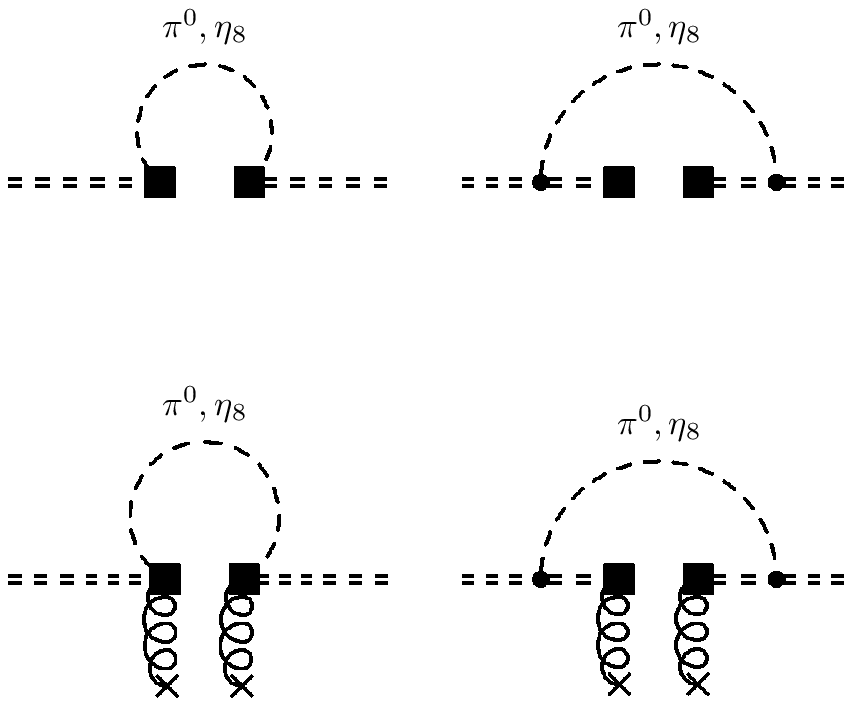,width=8cm,clip=true}
\caption{Chiral loop corrections to  the bag parameter.}
\label{fig:bagparam}
\end{center}
\end{figure}

 Now the bag parameter can be extracted and  may be written in the 
form:
\bea
\hat{B}_{B_q}=\fr{3}{4} \, \widetilde{b}
 \left[ 1 + \fr{1}{N_c}(1 - \delta_G^B)
+ \fr{\tau_b}{m_b} + 
\fr{\tau_\chi}{32 \pi^2 f^2} \right] \; \, ,
\label{Bhatform}
\eea
where 
\begin{equation}
\widetilde{b} \; = \; b(m_b) \, 
\left[ \fr{C_1-C_2}{(C_\gamma+C_v)^2}\right]_{\mu=\Lambda_\chi} \;  . 
\end{equation}

The soft gluonic  non-factorizable
 effects are given by
\begin{equation}
\delta_G^B \, = \, 
 \frac{N_c \gc}{32\pi^2 f^2 f_B^2} \, \frac{m}{M_B} \,
 \kappa_B \, \left[ \fr{ C_1}{C_1-C_2} \right]_{\mu=\Lambda_\chi} \; \ ,
\end{equation}
where $\kappa_B$ is a hadronic parameter (depending on
$m,f,\mu_G^2$ and $\ga$)
of order 2.             
Note that
 we are qualitatively in agreement  
with \cite{Mel}, where a negative contribution to the
 bag factor from gluon condensate effects  is found.
The formula (\ref{Bhatform}) is a generalization of a similar formula
found for $K- \overline{K}$ mixing \cite{BEFL}.

\begin{figure}[t]
\begin{center}
   \epsfig{file=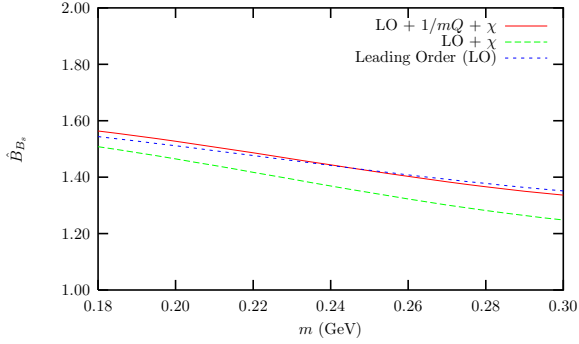,width=8cm}
\caption{The bag parameter $\hat{B}$ for $B_s$ as a function of $m$.}
\label{fig:Bbs}
\end{center}
\end{figure}

Numerically, $f$ and $f_B$ are of the same order of magnitude, and 
$\delta_G^B$ is therefore suppressed like $m/M_B$ compared to the corresponding
 quantity 
\begin{equation}
\delta_G^K \, = \, N_c 
 \frac{\gc}{32\pi^2 f^4} 
\end{equation}
for $K- \overline{K}$ mixing. However, one should note that $f_B$
scales as $1/\sqrt{M_B}$ within HQEFT, and therefore 
 $\delta_G^B$ is still formally of order  $(m_b)^0$.
 The quantity $\tau_b$  represents the $1/m_b$ corrections due to
the operators $X_i$. Furthermore, the quantity $\tau_\chi$ 
represents the chiral corrections
 to the bosonized versions of $Q_{1,2}$ \cite{ahjoeB} and corresponds to
 the diagrams in figure~\ref{fig:bagparam}.
The bag parameter $\hat{B}$ is plotted as function of $m$
in figure~\ref{fig:Bbs} for the case $B_s$.
From Table 1 and \cite{ahjoeB}
we observe that our results are numerically in agreement with 
recent lattice results \cite{latt}.

\section{The processes $B \rightarrow D \overline{D}$}

For these processes there are only two relevant four quark operators.
Within Heavy Quark Effective Theory (HQEFT)~\cite{neu},
the effective weak non-leptonic Lagrangian 
 can be evolved down to the scale
 $\mu \sim \Lambda_\chi \sim$1 GeV \cite{GKMWF}.  The $b$, $c$, and
$\overline{c}$ quarks are then treated within HQEFT. 
\begin{figure}[t]
\begin{center}
   \epsfig{file=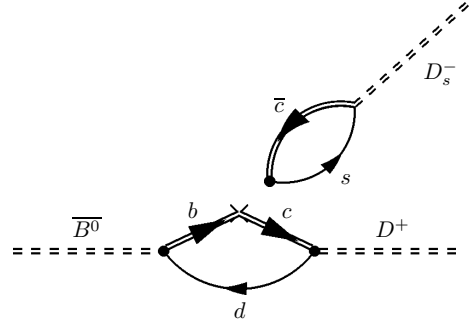,width=7cm}
\caption{Factorized contribution for 
$\overline{B^0}  \rightarrow D^+ D_s^-$
through the spectator mechanism, which does not exist for
the decay mode $\overline{B^0} \rightarrow D_s^+ D_s^-$
we consider in this paper.
 The double dashed lines represent heavy mesons, the double lines
 represent heavy quarks, and the single lines light quarks.}
\label{fig:bdd_fact}
\end{center}
\end{figure}
As an example of a typical {\it factorized} amplitude  we choose  
the case $\overline{B^0} \rightarrow D^+ D_s^-$
which is visualized in figure~\ref{fig:bdd_fact}:
\bea
&\,& A(\overline{B^0} \rightarrow D^+ D_s^-)_F =  \nonumber \\
  &-& \frac{G_F}{{\sqrt 2}}V_{cb} V_{cs}^*
 \, a_2 \,\zeta(\omega) f_D M_D 
\sqrt{M_B M_D} \; (\lambda + \omega) \; \, ,
\label{BDDF}
\eea
where $\zeta(\omega)$ is the Isgur-Wise function 
 for  the $\bar{B} \rightarrow D$  transition.
Here $\omega \equiv v \cdot v'= v \cdot \bar{v} = M_B/(2M_D)$ 
and $\lambda \equiv \bar{v} \cdot v' = (M_B^2/(2 M_D^2)-1)$, 
 where $v$, $v'$ and $\bar{v}$ are  the
 velocities of the heavy $b$-  $c$- and $\bar{c}$- quarks respectively.
 The Wilson coefficients $a_i$ contain short distance effects.
Numerically,
$a_1 \sim 10^{-1}$ and $a_2 \sim 1$ at the scale $\mu = m_b$.
For $\mu < m_c$ the  $a_i$'s are complex 
 and one has $|a_1| \simeq 0.4$ and $|a_2| \simeq 1.4$ at 
$\mu \sim \Lambda_\chi \sim$1 GeV \cite{GKMWF}.

\begin{figure}[t]
\begin{center}
   \epsfig{file=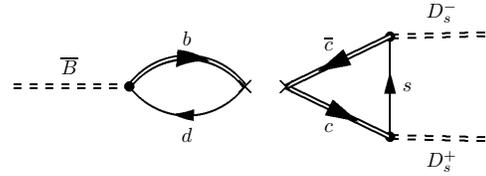,width=7cm}
\caption{Factorized contribution for 
$\overline{B^0}  \rightarrow D_s^+ D_s^-$
through the annihilation  mechanism, which give zero contributions if
both $D_s^+$ and $D_s^-$ are pseudo-scalars.}
 \label{fig:bdd_fact2}
\end{center}
\end{figure}

The factorized  amplitude for  $\overline{B^0} \rightarrow D_s^+ D_s^-$ 
 is visualized in figure~\ref{fig:bdd_fact2}.
Unless one or both of the $D$-mesons in the final state 
are vector mesons, this matrix
element is zero due to current conservation, which is
analogous to  the decay mode
$\overline{D^0}  \rightarrow K^0 \overline{K^0}$ \cite{EFZ}.

In the following we will consider explicitly the decay mode
$\overline{B^0} \rightarrow D_s^+  D_s^-$. The analysis of
$\overline{B^0_s} \rightarrow D^+  D^-$ proceed the same way.
To calculate the chiral loop amplitudes
 we need the factorized amplitudes for 
$\overline{B_s^{*0}} \rightarrow D_s^+ D^{*-}$ and
$\overline{B^0} \rightarrow D^{*+} D^{*-}$, which proceed through the
spectator mechanism as in  figure~\ref{fig:bdd_fact}.
\begin{figure}[t]
\begin{center}
   \epsfig{file=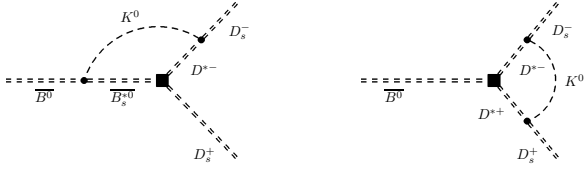,width=8cm}
\caption{Non-factorizable chiral loops for 
$\overline{B^0} \rightarrow D_s^+ D_s^-$.}
\label{fig:chiral1}
\end{center}
\end{figure}
We obtain the following chiral loop amplitude for the process
 $\overline{B^0} \rightarrow D_s^+ D_s^-$ 
from the figure~\ref{fig:chiral1}:
\bea
A(\overline{B^0} \rightarrow D_s^+ D_s^-)_\chi \, = \,  
\left(V_{cd}^*/V_{cs}^*\right)
 \, A(\overline{B_d^0} \rightarrow D_d^+ D_s^-)_F \cdot
\; R^\chi \; \, ,
\label{chiralR}
\eea
where the  factorized amplitude
for the process $\overline{B^0} \rightarrow D^+ D_s^-$ is
given in (\ref{BDDF}).

The quantity $R^\chi$ is a sum of contributions 
from the left and right
part of figure~\ref{fig:chiral1}, and proportional to 
$(m_K \ga/4\pi f)^2$ which is $1/N_c$ suppressed. Numerically,
\bea
R^\chi \;\simeq \; 0.12 - 0.26 i  \; .
\label{XNum}
\eea
The genuine non-factorizable part for 
 $\overline{B^0} \rightarrow D_s^+ D_s^-$
at quark level 
can, by means of Fierz transformations and the identity 
 (\ref{fierz}),   be written 
in terms of  coloured currents.

 The  left part in figure~\ref{fig:bdd_fact2} with gluon emission
 gives us the bosonized coloured current which is the same as for
 $B-\overline{B}$ mixing in eq. (\ref{1G}). 
\begin{figure}[t]
\begin{center}
   \epsfig{file=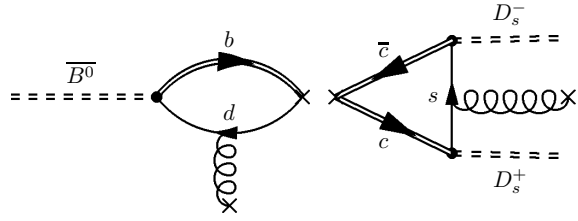,width=8cm}
\caption{Non-factorizable contribution for 
$\overline{B^0}  \rightarrow D_s^+ D_s^-$
through the annihilation mechanism with additional soft gluon emission.
 The wavy lines represent soft
gluons ending in vacuum to make gluon condensates.}
\label{fig:bdd_nfact2}
\end{center}
\end{figure}
For the creation of a $D \bar{D}$ pair in the right part of figure 
\ref{fig:bdd_fact2}, there is an  analogue of (\ref{1G}).
We find the gluon condensate contribution
for   $\overline{B^0} \rightarrow D_s^+ D_s^-$
within our model:
\bea
&\,& A(\overline{B^0} \rightarrow D_s^+ D_s^-)_G \; =  \nonumber \\  
  &-& \frac{G_F}{\sqrt{2}} V_{cb}V_{cq}^* \, a_2 \,  \gc \; 
\fr{(G_H \sqrt{M_B})^3}{384 m} \, F_G \; \; ,
\label{B2DD}
\eea
where $F_G$ is a dimensionless complex  function of $\lambda$.
The ratio between this  amplitude and the factorized one in
(\ref{BDDF}) scales as $M_D/(N_c M_B)$ times
hadronic parameters calculated within HL$\chi$QM. 
We define a quantity $R^G$ for the gluon
condensate amplitude  (\ref{B2DD}) analogously to  $R^\chi$ in (\ref{chiralR})
for chiral loops.
Numerically, we find that the ratio between the two amplitudes
 in (\ref{B2DD}) and (\ref{BDDF}) is
\bea
R_G \; \simeq \; 0.055  + 0.16 i  \; \; ,
\label{RNum}
\eea
which is of order one third of the chiral loop contribution in
eq. (\ref{XNum}).

Note that our non-factorizable amplitudes 
in (\ref{chiralR}) and (\ref{B2DD}) are proportional to the
numerically favourable Wilson coefficient $a_2$.

We find the branching ratios
\bea
\label{BR}
 BR(\overline{B^0_d} \to D_s^+ D_s^-) \simeq 
 7 \times 10^{-5}  \; \; , \\
 \quad
 BR(\overline{B^0_s} \to D^+ D^-) \simeq 1 \times 10^{-3} \; \; .
\eea
The difference between the two 
branching ratios is mainly due to the difference in
KM factor. 
For further details we refer to \cite{EFH}.

As mentioned above, the decay mode  
$\overline{D^0}  \rightarrow K^0 \overline{K^0}$ \cite{EFZ}
is analogues to $\overline{B^0_d} \to D_s^+ D_s^-$
in the sense that there are only non-factorizable contributions.
 The soft
gluonic effects are similar to that in 
figure~\ref{fig:bdd_nfact2}, but with $b$ replaced by $c$, which means
that the left part of the diagram can still be described within HL$\chi$QM.
In the right part of the diagram, $c \bar{c}$ has to be replaced by
$s \bar{s}$, and $s$ replaced by $d$.
 In addition there is a mass insertion for the light quark
part. However, the chiral loop diagrams are rather different in the two cases
because 
there is only light mesons in the final state for the mode 
 $\overline{D^0}  \rightarrow K^0 \overline{K^0}$.

\section{The process $B \rightarrow D \, \eta'$}
\begin{figure}[t]
\begin{center}
\epsfig{file=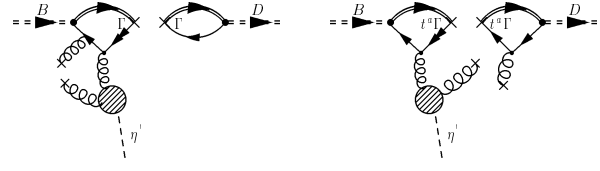,width=8cm}
\caption{Gluon condensate contributions to $B \rightarrow D \eta'$.}
\label{fig:BDeta}
\end{center}
\end{figure}

Within the HL$\chi$QM,  gluonic aspects of $\eta'$ may be treated \cite{EHP}.
Using Fierz transformations for the  four quark operators
for $b \rightarrow c d \bar{u}$, we obtain contributions corresponding
to  figure \ref{fig:BDeta}. 
In our approach two gluons are emitted from the light quark lines. One of
these (the virtual $g^*$)
attach to the  $\eta' g g^*$-vertex, and the other end in vacuum and
make a gluon condensate together 
with one of the other soft  gluons ($g$) from the
$\eta' g g^*$-vertex. This vertex  which can be written:
\begin{equation}
\label{etagg}
 -\frac{1}{2}F_{\eta^{'}g^{*}g}\delta^{ab}
\varepsilon^{\mu\nu\rho\sigma}\varepsilon_\mu^{*a}G_{\nu\rho}
^b(0) \, q_{\sigma} \; \, 
\end{equation}
where $G(0)$ is the soft gluon tensor, $\varepsilon$ is the
 polarization vector of the virtual gluon, and $q$ is the momentum of
 the $\eta'$.
 We have used existing parameterizations 
of the $\eta' g g^*$-vertex form factor $F$ in (\ref{etagg}) from the
 literature (at a
 scale of order 1 GeV they are numerically not very different).
We have assumed that the  current for $B\rightarrow g^*$ is related 
to the better known case $B\rightarrow \rho$:
\begin{eqnarray}
J^{\mu a}(B\rightarrow g^{*d}) 
= &\bra g^{*d}
|\overline{b}t^a\gamma^{\mu}(1-\gamma_5)d |B\ket\nonumber \\
=&\frac{\delta^{ad}}{2}\frac{g_s f_\rho}{m_\rho^2}\bra \rho
|\overline{b}\gamma^{\mu}(1-\gamma_5)d |B\ket 
\nonumber
\end{eqnarray}

It turns out that the ``factorizable'' diagram to the left in 
figure~\ref{fig:BDeta} can be neglected compared to the 
non-factorizable diagram to the right. Using the knowledge of the
$B \rightarrow \rho$ current matrix element,
we obtain the result \cite{EHP}
\begin{equation}
Br(B\rightarrow D\eta^{'}) \,=\, (2.2\, \pm\, 0.4)\,
\times \, 10^{-4} \; \, ,
\end{equation}
for $m$ in the range 230-270 MeV.

\section{The $\beta$ term for $D^*\to D\gamma$}
The chiral Lagrangian $\beta$-term has the form \cite{itchpt,IWSte}:
\begin{equation}
{\cal L}_{\beta}=\fr{e \beta}{4}Tr[\overline{H} \, H \, \sigma\cdot
F \, Q_q^\xi]\, .
\label{beta}
\end{equation}
Here $Q^\xi_q=(\xi^\dagger Q_q \xi+\xi Q_q \xi^\dagger)/2$, where $Q_q$ is the
$SU(3)$ charge matrix for light quarks, $Q_q = diag(-2/3,-1/3,-1/3)$,
 and $F$ is the electromagnetic field tensor.
The $\beta$ term can be calculated in 
HL$\chi$QM, by considering  diagrams which look like those in
figure~\ref{fig:va}, but with the vector and axial vector fields
${\cal V}_{\mu}$ or  ${\cal A}_\mu$ 
 replaced by a photon field tensor.
 To leading order, we obtained the following expression :
\begin{equation}\label{betaLO}
\beta_{LO} =
\fr{G_H^2 f^2}{2m^2}\left\{1+\fr{N_c m^2}{4\pi f^2}
-\left(\fr{56+3\pi}{576 f^2 m^2}\right)\gc\right\}\, .
\end{equation}

\begin{figure}
\epsfig{file=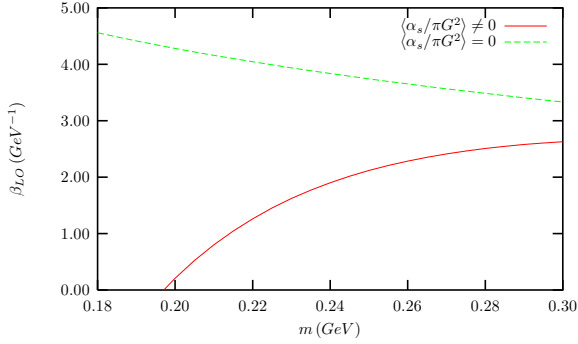,width=8cm}
\caption{$\beta_{LO}$ as a function of the constituent light quark mass
with and without gluon condensate.}
\label{fig:betap}
\end{figure}

As seen from figure~\ref{fig:betap}, $\beta$ depends strongly on the
constituent light quark mass $m$ because
 there is a  
partial   cancellation between large terms in
(\ref{betaLO}).
One may hope that $1/m_c$ corrections might help to stabilize the
result , but this is not the case. In fact,
 $1/m_c$ corrections
do not play a significant role for $m >$ 230 MeV.
(For smaller  $m$ they even pull in the wrong direction compared to
 the experimental value of order 2-3 GeV$^{-1}$ \cite{Beta}).
To obtain a value close to the experimental value for $\beta$,
 we need a  value
for $m$ higher than used in \cite{ahjoeB,AHJOE}.
Choosing $m$ in the range 250-300 MeV we find 
 $\beta=(2.5 \pm 0.6)$ GeV$^{-1}$ to be compared with
$\beta= (2.7 \pm 0.20)$ GeV$^{-1}$ extracted from experiment.
For further details we refer to \cite{Beta}.

\section{Conclusion}

In \cite{AHJOE} we have constructed a heavy-light chiral quark
model including soft gluonic effects and chiral loops.
The model describes the heavy-light
sector reasonably well\cite{ahjoeB,EFH,EHP,EFZ,Beta,AHJOE}.
 There is, however a difference compared to
the pure light sector where $f_\pi$ is  precisely known.
If $f_D$ and $f_B$ had been more 
precisely known, we would have  used them as numerical
input (together with $\ga$) to fix  $m$ and $\qc$ within the model.
Instead  we have used typical values of $m$ and $\qc$ 
(and $\ga$) as input, while $f_D$ and $f_B$ become output. 
(For a very recent review on numerical values for $f_D$ and $f_B$, 
see \cite{becir})

The value of $\beta$ turned out to be rather unstable \cite{Beta}
for the values
of $m$ and $\qc$ used in \cite{ahjoeB,AHJOE}. Higher values of $m$
are needed to obtain an acceptable $\beta$ in agreement with experiment. 
 However, this will lead to 
values of $f_B$ and
$f_D$ which are too small .
 This can be compensated by  using a higher value of the quark
condensate $\qc$.
 This is acceptable because  our model dependent quantities
 $\qc$ and  $\gc$ are not necessarily 
 exactly those obtained
in QCD sum rules.
\begin{center}
\begin{table}[ph]
\caption{Numerical values for the $B$-sector.}
\label{tab:test}
{\footnotesize
\begin{center}
\begin{tabular}{|c|r|r|}
\hline
{} &{} &{} \\[-1.5ex]
{} & Input values I & Input values II  \\[1ex]
\hline
{} &{} &{} \\[-1.5ex]
$G_H$ & $\quad(8.3\pm 0.7)\,$ GeV$^{-1/2}$& $\quad(7.2\pm 0.5)\,$
GeV$^{-1/2}$ \\[1ex]
$\gc^{1/4}$ & $\quad(300\pm 25)\,$ MeV & $\quad(340\pm 20)\,$ MeV \\[1ex]
$f_B$ & $\quad(190\pm 50)\,$ MeV  & $\quad(185\pm 30)\,$ MeV\\[1ex]
$f_{B_s}$ & $\quad(210\pm 70)\,$ MeV & $\quad(215\pm 45)\,$ MeV \\[1ex]
$f_{B_s}/f_B$ & $\quad 1.14\pm 0.07$ & $\quad 1.22\pm 0.02$ \\[1ex]
$\hat{B}_{B_d}$ & $\quad 1.51\pm 0.09$ & $\quad 1.52\pm 0.07$\\ [1ex]
$\hat{B}_{B_s}$ & $\quad 1.4\pm 0.1$   & $\quad 1.4\pm 0.1 $\\[1ex]
$\xi $          & $\quad 1.08\pm 0.07$ & $\quad 1.16\pm 0.04$ \\[1ex]
\hline
\end{tabular}
\end{center}
}
\vspace*{-13pt}
\end{table}
\end{center}
In Table 1 we have given our numerical values for a few important
 quantities in the $B$-sector. We have 
considered two sets of input. The first one is I: 
$m$ = 190 to 250 MeV and 
$-\qc^{1/3}$ = 230 to 250 MeV \cite{ahjoeB,Beta}. 
The second one is II:  $m$ = 250 to 300 MeV 
and $-\qc^{1/3}$ = 250 to 270 MeV \cite{Beta}. In both cases, $\ga=0.59$.
For further details we refer to \cite{ahjoeB,Beta,AHJOE}.
Note that the quantity $\xi$ in the table is defined as
$\xi=\left(f_{B_s}\sqrt{\hat{B}_{B_s}}\right)/
\left(f_{B_d}\sqrt{\hat{B}_{B_d}}\right)$
as usual.
We observe that $\hat{B}$ is very stable with respect to variations
in the input parameters.

\begin{figure}
\epsfig{file=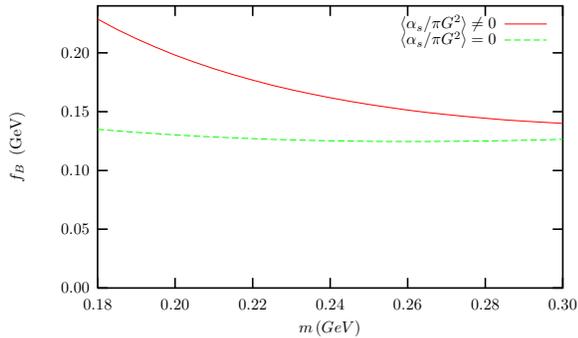,width=8cm}
\caption{\small{$f_B$ as a function of the constituent  light quark mass
with and without gluon condensate.}}
\label{fig:fB01G}
\end{figure}

Our model \cite{AHJOE} is different from \cite{barhi,CQM} in the sense that
we include the (phenomenological) gluon condensate. The figures
\ref{fig:betap} and \ref{fig:fB01G} illustrates the importance of the
gluon condensate at different values of $m$. In figure~\ref{fig:fB01G}
the  curves will be lifted for the quark condensate value II.

\section*{Acknowledgments}
I thank the organizers for warm hospitality.
I also thank my collaborators S.Fajfer, A. Hiorth, A. Polosa and J. Zupan.
This work is supported in part by  the Norwegian research council and  by
 the European Union RTN
network, Contract No. HPRN-CT-2002-00311  (EURIDICE).


\end{document}